\documentclass[conference]{IEEEtran}
\IEEEoverridecommandlockouts
\usepackage{cite}
\usepackage{amsmath,amssymb,amsfonts}
\usepackage{algorithmic}
\usepackage{graphicx}
\usepackage{textcomp}
\usepackage{xcolor}
\def\BibTeX{{\rm B\kern-.05em{\sc i\kern-.025em b}\kern-.08em
    T\kern-.1667em\lower.7ex\hbox{E}\kern-.125emX}}
\begin{document}

\title{Advanced Audio Aid for Blind People 
}

\author{\IEEEauthorblockN{1\textsuperscript{st} Savera Sarwar}
\IEEEauthorblockA{\textit{Department of Electronics Engineering} \\
\textit{Quaid-e-Awam University of Engineering, Science and Technology}\\
Larkana, Pakistan \\
saverasarwar1@gmail.com}
\and
 \IEEEauthorblockN{2\textsuperscript{nd} Muhammad Turab                 }
 \IEEEauthorblockA{\textit{Department of Computer Systems Engineering} \\
 \textit{Mehran University of Engineering and Technology}\\ 
 Jamshoro, Pakistan \\
 turabbajeer202@gmail.com}
 \and
 \IEEEauthorblockN{3\textsuperscript{rd} Danish Channa                 }
 \IEEEauthorblockA{\textit{Department of Electronics Engineering} \\
 \textit{Quaid-e-Awam University of Engineering, Science and Technology}\\ 
 Larkana, Pakistan \\
 alid07818@gmail.com}
 \and
 \IEEEauthorblockN{4\textsuperscript{th} Aisha Chandio}
 \IEEEauthorblockA{\textit{Department of  Computer Science} \\
 \textit{Quaid-e-Awam University of Engineering, Science and Technology}\\
 Nawabshah, Pakistan \\
 aishabatool5512@gmail.com}
 \and

 \IEEEauthorblockN{5\textsuperscript{th} M. Uzair Sohu}
 \IEEEauthorblockA{\textit{Department of Electrical Engineering} \\
 \textit{Quaid-e-Awam University of Engineering, Science and Technology}\\ 
 Larkana, Pakistan \\
 Mohammeduzairsohu123@gmail.com}
 
 \and
 \IEEEauthorblockN{6\textsuperscript{th} Vikram Kumar}
 \IEEEauthorblockA{\textit{Department of Electrical Engineering} \\
 \textit{Quaid-e-Awam University of Engineering, Science and Technology}\\ 
 Larkana, Pakistan \\
 vk2391080@gmail.com}

}

\maketitle

\begin{abstract}

One of the most important senses in human life is vision,
without it one’s life is totally filled with darkness. According to WHO
globally millions of people are visually impaired estimated there are
285 million, of whom some millions are blind. Unfortunately, there
are around 2.4 million people are blind in our beloved country
Pakistan. Human are a crucial part of society and the blind community
is a main part of society. The technologies are grown so far to make
the life of humans easier more comfortable and more reliable for.
However, this disability of the blind community would reduce their
chance of using such innovative products. Therefore, the visually
impaired community believe that they are burden to other societies
and they do not capture in normal activities separates the blind
people from society and because of this believe did not participate in
the normally tasks of society . The visual impair people mainly face
most of the problems in this real-time The aim of this work is to turn
the real time world into an audio world by telling blind person about
the objects in their way and can read printed text. This will enable
blind persons to identify the things and read the text without any
external help just by using the object detection~\cite{chandio2022precise} and reading system in
real time. Objective of this work: i) Object detection ii) Read printed text, using state-of-the-art (SOTA) technology.
\end{abstract}

\begin{IEEEkeywords}
Audio classification, Aid blind people, Text detection, object detection, Optical character recognition (OCR)
\end{IEEEkeywords}

\section{Introduction}
According to the (WHO) statistics~\cite{WHO22}, the most common disabilities worldwide are blindness and loss of vision. More than 100 million peoples have vision blindness and it is getting more common day by day. At least half of a billion had these cases, vision impairment is always to be unanswered and got propriety to live in this society. This at least millions of
people include those with small distance vision impairment or blindness due to unanswered refractive error, diabetic retinopathy, trachoma, cataract, corneal opacities as well as near vision impairment caused by unanswered presbyopia. Population growth and aging are the most common reasons that more people sustain vision impairment nowadays more and more people are getting into it by the time. In Pakistan, more than 200+ million population lives with blindness and cases are increasing over time, as last year’s (NCPB) shows. blind has a much impact on vision impairment peoples, including the ability to steer, read and steer the circumstances of the environment independently and easily. A white cane is used by blind people that does not provide more information and details about the environment. Due to this, they are facing lots of issues in this society, including difficulty in walking and understanding the obstacles most of them are using Trained guide dogs that can assist the blind. Which are very expensive. On one side from the achievement of biotechnologies and machines and madicans to find a proper solution to visually impaired people's problems other technologies like electronic devices and computers can provide finer tools to improve their quality of life in society and give them better hand to live much better in this society. Most of the projects have come up recently. Lots of the devices had sensors to understand the environment and as result, it gives voice or sound through headphones. The quality of Audio and speed time response affect the reliability of these devices. Many devices, in the market, do not contain a real-time reading assistant and had a poor UI, are expensive, less portable, and lack hands-free access. These devices are not that popular among the visually impaired and need to have an improvement in design, performance (speed of response), and reliability by using both indoors and most importantly outdoors. AI, Machine
Learning, and Image and Text Recognition is recently been widely used in devices to assist blind people in daily life. Our whole project is based on AI (Computer vision) deep learning and image processing techniques.  It takes input through the
camera process and turns the worldly The main goal of the project is to provide visually impaired people with high accuracy, the best performance results, and a realistic choice to make the world a happy place for them.
urrently deep
learning (DL) algorithms have shown state-of-the-art (SOTA) performance in many domains, i.e Text, image~\cite{krizhevsky2012imagenet, kumar2021binary, kumar2021class, kumarstride, kumarforged,kumardetection} and audio data ~\cite{chandio2021audd,park2020search,kumar2020intra,turab2022investigating}. Exploiting these algorithms have been a challenging for real time application. These SOTA DL algorithms can be beneficial for blind people to guide them about object in their way. To do so, we propose approach that can turn audio world into real world for blind people.
\\
Our work contribution is summarized as of the following:
\begin{itemize}
    \item To the best of our knowledge, we are the first to turning the audio world into real world for blind people. 
    \item We propose novel approach to help blind people using SOTA DL algorithms.
    \item To check performance of DL algorithms, we show the results of text and image algorithms.
\end{itemize}
Rest of our paper is organized as in sections, \ref{LR} discusses literature review, \ref{impl} explains our proposed methodology, \ref{results} presents result of SOTA DL methods and finally \ref{conclude} conclude the work.

\section{Literature Review}\label{LR}
Object detection and text recognition have been challenges for the computer fields. There are a bunch of research papers that have talked about projects and their drawbacks and also discussed different methods to be used for the project. In this
work, the project is for completely blind people by using visual aid system. Because of its reasonable cost, small size, in
For this project, we have used Raspberry Pi three with a Model of type B+ to signify the scalability of the prototype. The design includes a camera and some sensors for obstacle avoidance and image processing for object detection which will come in front of blind people. The space between the blind and the hurdle is processed by the camera as well as ultrasonic sensors. The system contains a reading assistant, in the form of the image-to-text converter, which is followed by auditory feedback \cite{b2} although this system is providing three features but Raspberry Pi 3+ Ram is not enough to run these all with good speed. Good speed is very essential when using deep learning techniques. Raspberry PI 4 B is used in our project to give good speed. The pi camera is very sensitive to how a blind person got to manage that's why in our project we have used a webcam that can be easily used. In this prototype, we
have used Raspberry Pi Camera and which will help capture the picture, and that picture is converted into a scanned image then use the software Imagemagick for the further process. This Imagemagick software will provide us a scanned image then that scanned image is given to the Tesseract OCR (Optical Character Recognition) as an input, and OCR software convert the image into the text \cite{b21}. To convert the text into speech we use TTS (Text to Speech) engine \cite{b10} Taking the
pictures and processing first through Imagemagick software and then the OCR is a little bit time consuming. Several types purposed in recent studies \cite{b16, b17} of wearable. The quality of the audio signal, convey in real-time, affects the reliability of these devices. The system of assistant which is based on vision uses different types of cameras, such as stereo cameras, mono cameras, and RGB-D cameras, to capture images from the real-world environment, and use computer algorithms based on vision such as segmentation of the image, edge detection, and image filtering to detect objects \cite{b1}. For
object detection, artificial neurons are used in deep neural networks they are similar to humans composed of neurons and process forwarding. Most People have tried hard to solve this problem, but they are able to achieve 65\% of accuracy yet. It is so hard for machines to understand and classify objects like humans. Each part of the object in an Image or scene is recognized by the computer/software called object detection \cite{b14}. Object detection methods based on deep learning approach, such as Region-based Convolutions Neural Networks (R-CNN) for object detection, Spatial Pyramid Pooling Networks (SPPNet) give a variable size to the input image for object detection, fast R-CNN, faster R-CNN, (R-FCN), (FPN), and (YOLO) show us advantages \cite{b6, b7, b8, b9, b10, b11}. The YOLO-V3 network is the part of the YOLO which uses the DarkNet-53 as the model of the backbone. Compared with the Faster R-CNN network, the detection issue into regression Issues are transformed by the YOLO network. It does not require a Region Proposal Network, and it generates bounding box coordinates and probabilities of each class directly through regression \cite{b13}. Yolov3 is now an old version of Yo lo new versions of yolov4 even yolov5 are available. Yolov5 is very fast \cite{b9}, and lightweight than Yolo's previous versions and that’s why we deployed it in our project. It has very high accuracy as compared to other object detection models. We have tried Convolutions Neural Network to apply
on a Raspberry Pi. Raspberry Pi is a small computer that is relatively inexpensive and can be run on Linux-based operating systems as open source. Raspberry Pi works on python programming thus it makes it easier for deep learning programs which is built on Python \cite{b12} this system uses tensor flow using python language but for purposes, we have used PyTorch framework which is simple and also Open CV library. n this research OCR and Text-to-speech are used to convert these images in audio output (speech) \cite{b4, b8} synthesis. Our project is focused on providing Visually Impaired people with a system having two features object detection and reader and that can be carried around easily.

\section{ IMPLEMENTATION AND METHODLOGY }\label{impl}
In this chapter we will discuss the different methodologies to implement the purposed project. As it is hardware plus software project .In the previous chapter we see the hardware components and now will do work on software which starts from installation of OS and ends at the implementation of object detection and text detection recognition.

\begin{figure}[tp!]
\centering 
\includegraphics{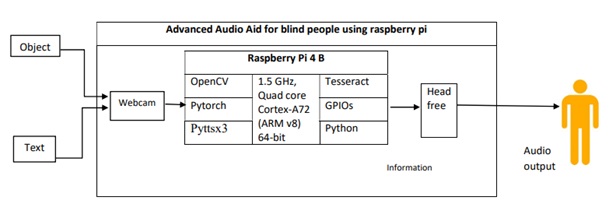}
\caption{ Advanced Audio system. Figure taken from}
\label{fig}
\end{figure}

\subsubsection{OPERATING SYSTEM}
In the project, we installed Raspberry Pi 32-bit OS and followed these steps. The Raspberry Pi Raspbian, a Debian-based(32-bit) Linux distribution, and Ubuntu, First, download the Raspberry Pi OS image and Now put the SD card reader in the PC, then open the Imager and select the OS (Raspberry Pi OS (32-bit)) and SD card to write the OS. And Connect 
\subsection{OpenCV} 
 OpenCV is in our project it is very important to install it for object detection and text recognition.

\begin{itemize}
\item Step1: Open the command line and update the system by following commands. 
\item Step2: Increase the GPU memory, by using these commands. On a Raspberry Pi 3, the GPU will take 128 MB. first, it is necessary to change it. 
\item Step 3: Install all Dependencies 
\end{itemize}
\subsection{Object detection}

Object detection is a technique of computer vision that permit us to recognize and locate objects in an image or video. We have used YOLO, It is a novel convolutional neural network (CNN) that detects objects in real-time with great accuracy. It processes the picture and then separates it into a boundary box for each component. OLOv5 is a family of
object detection architectures and COCO dataset is used to pretrained the models.

\begin{figure}[tp!]
\centerline{\includegraphics{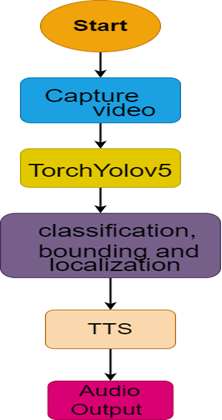}}
\caption{Yolov5 flow chart.}
\label{fig_yolo}
\end{figure}

\subsection{Text Reader} 

The reader reads the text from the image first applying image processing techniques then applying OCR. the unwanted noise is removed by using image processing techniques like gray scaling, blur, and threshold. OCR technology allows the
scanned images of printed text or symbols to be converted into text or information that can be easily understood or edited by using the program. In our system, we have used the TESSERACT library for OCR technology. Tesseract is an optical character recognition engine. For the text search and its recognition of images, OCR uses artificial intelligence for it. Tesseract is finding templates in pixels, letters, words, and sentences. To convert the data into audio Text-to-speech recognition library is used.

\begin{figure}[tp!]
\centerline{\includegraphics{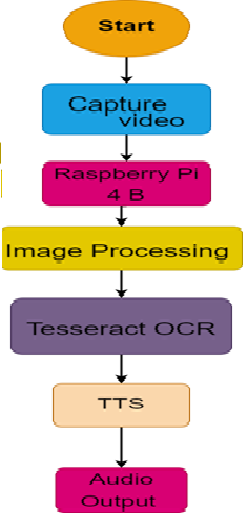}}
\caption{Text reader.}
\label{fig_text_reader}
\end{figure}
\subsection{Convert text to voice}

we have used pyttsx3 is a text-to-speech conversion library in Python which is used to convert the text into voice.

\begin{figure}[tp!]
\centerline{\includegraphics{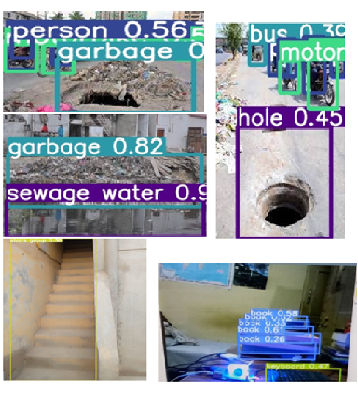}}
\caption{Detection in different scenarios .}
\label{fig1}
\end{figure}
\begin{figure}[tp!]
\centerline{\includegraphics[width=0.5\textwidth]{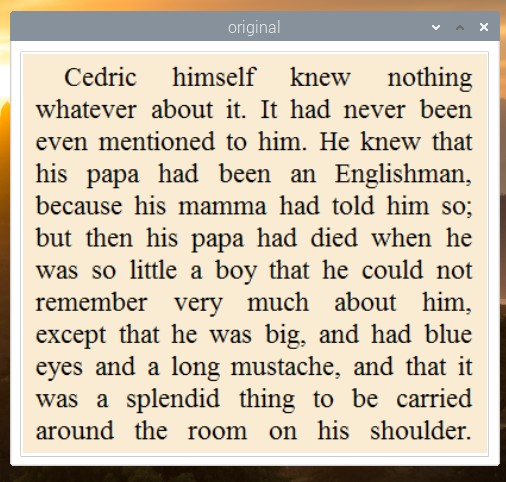}}
\caption{Color image .}
\label{fig2}
\end{figure}
\begin{figure}[tp!]
\centerline{\includegraphics[width=0.5\textwidth]{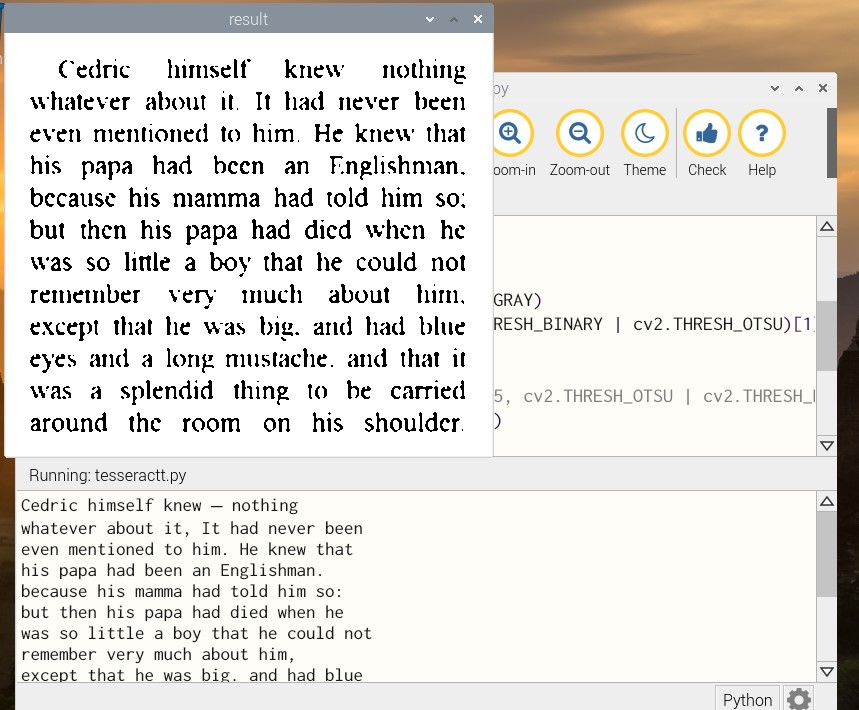}}
\caption{Colored image output with text .}
\label{fig3}
\end{figure}

\begin{figure}[htbp]
\centerline{\includegraphics{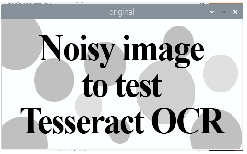}}
\caption{ Noisy image. Image taken from \cite{b22}}
\label{fig4}
\end{figure}

\section{  Result: }\label{results}
{
The offered methodology is tested by a sample of some text images and object images, state-of-the-art models have shown quite good results and simulation of text and image object is shown in figures \ref{fig1}, \ref{fig2}, \ref{fig3} and \ref{fig4}. In figure \ref{fig1}, the objection algorithm has detected different image objects like ladders, dangerous holes, garbage, etc. In figure~\ref{fig2}, the color image is passed to OCR for text extraction, it successfully extracted the text irrespective of color, as shown in the figure~\ref{fig3}. Even if we tried noisy image \ref{fig4}, OCR was able to extract text with a very good results. So it was shown, these algorithms of image and text extraction are quite helpful for our purpose.}

\section*{Conclusion}\label{conclude}
We have successfully implemented an “Advanced Audio Aid for Blind people using Raspberry Pi”. The results of the experiment are verified successfully and the hardware output is also verified. It is capable of reading text and identifying objects including stairs, path holes, sewage water, etc. The device output is in the form of voice so, it can be easily heard by visually impaired people. This system is an efficient device as well as economically helpful for blind people. This
device is useful for daily life and in blind schools, and colleges and can be also used as an application of artificial intelligence. illiterate people can also be helpful by this device and this device is compact in size and very useful to the society. In the future we try to increase the accuracy of object detection by increasing the number of images of the dataset and with this sort of identification and localization, object detection will use to count objects in a scene and determine and track their precise locations, all while accurately labeling them. This will help in navigation and prevention. We also improve the reader by adding some new techniques like OpenCV EAST detection which we could not use because of short time of frame.


\begin{thebibliography}{00}

\bibitem{WHO22} WHO, 'Blindness and vision impairment'  (2021), https://www.who.int/news-room/fact-sheets/detail/blindness-and-visual-impairment, Accessed{25/06/2022}

\bibitem{chandio2021audd}Chandio, A., Shen, Y., Bendechache, M., Inayat, I. \& Kumar, T. AUDD: audio Urdu digits dataset for automatic audio Urdu digit recognition. {\em Applied Sciences}. \textbf{11}, 8842 (2021)
\bibitem{park2020search}Park, J., Kumar, T. \& Bae, S. Search for optimal data augmentation policy for environmental sound classification with deep neural networks. {\em Journal Of Broadcast Engineering}. \textbf{25}, 854-860 (2020)
\bibitem{kumar2020intra}Kumar, T., Park, J. \& Bae, S. Intra-Class Random Erasing (ICRE) augmentation for audio classification. {\em Proceedings Of The Korean Society Of Broadcast Engineers Conference}. pp. 244-247 (2020)
\bibitem{turab2022investigating}Turab, M., Kumar, T., Bendechache, M. \& Saber, T. Investigating Multi-Feature Selection and Ensembling for Audio Classification. {\em ArXiv Preprint ArXiv:2206.07511}. (2022)

\bibitem{krizhevsky2012imagenet}Krizhevsky, A., Sutskever, I. \& Hinton, G. Imagenet classification with deep convolutional neural networks. {\em Advances In Neural Information Processing Systems}. \textbf{25} (2012)
\bibitem{kumar2021binary}Kumar, T., Park, J., Ali, M., Uddin, A., Ko, J. \& Bae, S. Binary-classifiers-enabled filters for semi-supervised learning. {\em IEEE Access}. \textbf{9} pp. 167663-167673 (2021)
\bibitem{kumar2021class}Kumar, T., Park, J., Ali, M., Uddin, A. \& Bae, S. Class Specific Autoencoders Enhance Sample Diversity. {\em Journal Of Broadcast Engineering}. \textbf{26}, 844-854 (2021)
\bibitem{kumarstride}Kumar, T., Brennan, R. \& Bendechache, M. STRIDE RANDOM ERASING AUGMENTATION. 
\bibitem{kumarforged}Kumar, T., Turab, M., Talpur, S., Brennan, R. \& Bendechache, M. FORGED CHARACTER DETECTION DATASETS: PASSPORTS, DRIVING LICENCES AND VISA STICKERS. 

\bibitem{kumardetection}Kumar, Teerath and Turab, Muhammad and Talpur, Shahnawaz and Brennan, Rob and Bendechache, Malika. DETECTION DATASETS: FORGED CHARACTERS FOR PASSPORT AND DRIVING LICENCE.


\bibitem{b1} Jinqiang Bai, Zhaoxiang Liu, Yimin Lin, Ye Li, Shiguo Lian, Dijun Liu. (2019). “Wearable travel aid for environment per caption and navigation of visually impaired people”.

\bibitem{b2} Muiz Ahmed Khan, Pias Paul, Mahmudur Rashid, Mainul Hossain, and Md Atiqur Rahm. (2020). “An AI Based Visual Aid with Integrated Reading Assistant for The Completely Blind”. IEEE Transactions on Human-Machine Systems. pp 1-11.
\bibitem{b3} Blindness and vision impairment, World Health Organization, Geneva, Switzerland, 14 Oct. 2021. [Online]. https://www.who.int/news-room/fact-sheets/detail/blindness-and-visual-impairment
\bibitem{b4} Akila, I S; Akshaya, B; Deepthi, S; Sivadharshini, P (2018). “A Text Reader for the Visually Impaired using Raspberry Pi”. In Proceedings of the Second International Conference on Computing Methodologies and Communication (ICCMC 2018) IEEE Conference, pp 778-782.
\bibitem{b5}Kalyani, A., B. Premalatha, and K.R. Kiran, 2018. “Real Time Emotion Recognition from Facial Images using Raspberry Pi.”  International Journal of Advanced Technology.
\bibitem{b6} Fang, Wei; Wang, Lin; Ren, Peiming (2020). “Tinier-YOLO: A Real-Time Object Detection Method for Constrained Environments”. IEEE Access, pp. 1935-1944.
\bibitem{b7} Liu, Chengji; Tao, Yufan; Liang, Jiawei; Li, Kai; Chen, Yihang (2018). “Object Detection Based on YOLO Network”. IEEE 4th Information Technology and Mechatronics Engineering Conference (ITOEC), pp. 799–803.
\bibitem{b8} Anush Goel1, Akash Sehrawat2, Ankush Patil3, Prashant Chougule4, Supriya Khatavkar5 (2018).” Raspberry Pi Based Reader for Blind People”. International Research Journal of Engineering and Technology (IRJET).
\bibitem{b9} Xingkui Zhu, Shuchang Lyu, Xu Wang Qi Zhao (2021).” TPH-YOLOv5: Improved YOLOv5 Based on Transformer Prediction Head for Object Detection on Drone-captured Scenarios”. IEEE Xplore.
\bibitem{b10} Mainkar, Vaibhav V.; Bagayatkar, Tejashree U.; Shetye, Siddhesh K.; Tamhankar, Hrushikesh R.; Jadhav, Rahul G.; Tendolkar, Rahul S. (2020). “Raspberry pi based Intelligent Reader for Visually Impaired Persons”. IEEE 2nd International Conference on Innovative Mechanisms for Industry Applications (ICIMIA) - Bangalore, India, pp. 323–326.
\bibitem{b11} Alahmari, Saeed S.; Goldgof, Dmitry B.; Mouton, Peter R.; Hall, Lawrence O. (2020). “Challenges for the Repeatability of Deep Learning Models”. IEEE Access, pp. 211860–211868.
\bibitem{b12}Param Popat, Prasham Sheth, Swati Jain (2019).” Animal/Object Identification Using Deep Learning on Raspberry Pi”. S. C. Satapathy and A. Joshi (eds.), Information and Communication Technology for Intelligent Systems, Smart Innovation, Systems and Technologies 106.
\bibitem{b13}Yunong Tian, Guodong Yang, Zhe Wang, Hao Wang, En Li, Zize Liang (2019). “Apple detection during different growth stages in orchards using the improved YOLO-V3 model”. Computers and Electronics in Agriculture 157, pp. 417–426.
\bibitem{b14}Ashwani Kumar, S S Sai Satyanarayana Reddy, Vivek Kulkarni (2019).” An Object Detection Technique for Blind People in Real-Time Using Deep Neural Network”. 2019 Fifth International Conference on Image Information Processing (ICIIP), IEEE Xplore.
\bibitem{b15} Zhao, Zhong-Qiu; Zheng, Peng; Xu, Shou-Tao; Wu, Xindong (2019). “Object Detection with Deep Learning: A Review”. IEEE Transactions on Neural Networks and Learning Systems, pp. 1–21.
\bibitem{b16}J. Bai, S. Lian, Z. Liu, K. Wang, and D. Liu, (2017). “Smart guiding glasses for visually impaired people in indoor environment,” IEEE Trans. Consum. Electron. vol. 63, no. 3, pp. 258–266,
\bibitem{b17}J. Bai, S. Lian, Z. Liu, K. Wang, and D. Liu, “Virtual-blind-road following based wearable navigation device for blind people,” IEEE Trans. Consum. Electron, vol. 64, no. 1, pp. 136–143, Feb. 2018.
\bibitem{b18}In Pakistan, timely help could save vision of 85\% blind women March 06, 2021.[online] https://www.arabnews.pk/node/1820356/pakistan
\bibitem{b19}What is computer vision [online] https://www.ibm.com/topics/computer-vision
\bibitem{b20} Adwitiya Arora, Atul Grover, Raksha Chugh, S. Sofana Reka (2019).” Real Time Multi Object Detection for Blind Using Single Shot Multibox Detector”.
\bibitem{b21}Mahalakshmi, V. \& Bennet, M. \& Hemaladha, R. \& Jenitta, J.\& Vijayabharathi, K.. (2018). Implementation of OCR using raspberry PI for visually impaired person. International Journal of Pure and Applied Mathematics. 119. 111-117.

\bibitem{chandio2022precise}Chandio, A., Gui, G., Kumar, T., Ullah, I., Ranjbarzadeh, R., Roy, A., Hussain, A. \& Shen, Y. Precise Single-stage Detector. {\em ArXiv Preprint ArXiv:2210.04252}. (2022)


\bibitem{b22} Adrian Rosebrock , (2017), Using Tesseract OCR with Python,  March 06, 2021.[online]https://pyimagesearch.com/2017/07/10/using-tesseract-ocr-python/
\end{thebibliography}
\end{document}